\documentclass{article}[11pt,A4paper]

\usepackage{graphicx}
\usepackage{caption}
\usepackage{subcaption}
\usepackage[OT1]{fontenc}
\usepackage[latin1]{inputenc}
\usepackage[english]{babel}
\usepackage{amssymb,amsmath}

\makeatletter
\renewcommand{\fnum@figure}{\textbf{Figure~\thefigure}}
\makeatother

\begin{document}

\title{The YAPFI phase-field implementation}
\author{Henrik Larsson\footnote{\noindent e-mail:~hlarsso@kth.se}}

\maketitle

\begin{center}
Unit of Structures, Dept Materials Science and Engineering\\ KTH, SE-10044 Stockholm, Sweden
\end{center}

\begin{center}
\textbf{Abstract}
\end{center}

A fully implicit phase field model has been implemented in Fortran for 1-3D simulations. It is intended for simulations of diffusion and diffusion controlled transformations. Gibbs energy contributions from both gradients in concentration and gradients in phase field variables are included. Orientation dependent interfacial energy is also supported. The implementation supports  the use of Calphad type databases as well as analytical Gibbs energy expressions, though in the latter case phases may not have internal degrees of freedom 

\newpage

\tableofcontents

\section{A short introduction to the phase-field method}

A large number of introductory and review articles on the phase-field method have been written, see e.g.~Refs.~\cite{2002che}\cite{2008sin}\cite{2008moe}. This section only provide a very short background to the phase-field method.

The notation and syntax in this short introduction may differ slightly from that used in the references.

%--------------
\subsection{Spinodal decomposition}

\subsubsection{Background to spinodal decomposition}

Consider a binary system A-B consisting of a single regular solution phase $\alpha$. The molar Gibbs energy of $\alpha$ is given by

\begin{equation}
G_m^\alpha =x_A\,^\circ G_A^\alpha+x_B\,^\circ G_B^\alpha+RT\left(x_A\ln x_A+x_B\ln x_B\right)+x_Ax_BL_{AB}^\alpha
\end{equation}

If the regular solution parameter $L_{AB}^\alpha >0$ there will be a miscibility gap; inside the miscibility gap the equilibrium state is $\alpha^\prime +\alpha^{\prime\prime}$ where  $\alpha^\prime$ and $\alpha^{\prime\prime}$ have the same structure but different compositions. 

\begin{figure}
\begin{center}
\includegraphics[width=0.9\textwidth]{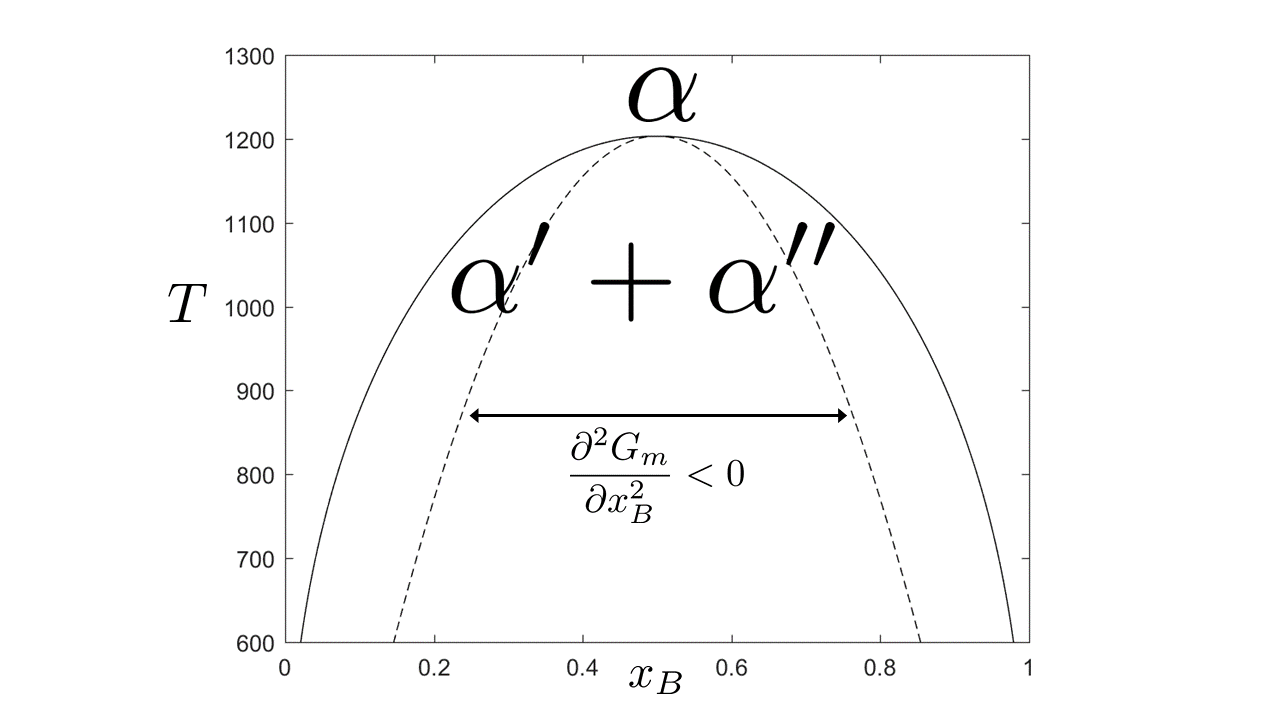}
\end{center}
\caption{Phase diagram with a miscibility gap. The dashed line show the extent of the spinodal.}\label{fig:miscibility_gap}
\end{figure}

Inside the miscibility gap there will be a so-called spinodal. For a given temperature and constant pressure, the limits of the spinodal are given by

\begin{equation}
\frac{\partial^2 G_m^\alpha}{\partial x_B^2}=0
\end{equation}

Inside the spinodal we have

\begin{equation}
\frac{\partial^2 G_m^\alpha}{\partial x_B^2}<0
\end{equation}

Fig.~\ref{fig:miscibility_gap} show a phase diagram with a miscibility gap where the extent of the spinodal is indicated with the dashed line.
 
For this single phase binary system, consider a homogeneous volume element and the exchange of some amount of $B$ for $A$  from one half of the volume element to the other. The resulting change in $G_m$ is

\begin{equation}
\begin{split}
\Delta G_m &=\frac{1}{2}\left[G_m\left(x_B+\Delta x_B\right)+G_m\left(x_B-\Delta x_B\right)-2G_m\left(x_B\right) \right]\\
& =\frac{1}{2}\left[\frac{G_m\left(x_B+\Delta x_B\right)-G_m\left(x_B\right)}{\Delta x_B}-\frac{G_m\left(x_B\right)-G_m\left(x_B-\Delta x_B\right)}{\Delta x_B}\right]\Delta x_B\\
& \simeq\frac{1}{2}\left[\frac{\frac{\partial G_m}{\partial x_B}\Big\vert_{x+\frac{1}{2}\Delta x_B} -\frac{\partial G_m}{\partial x_B}\Big\vert_{x-\frac{1}{2}\Delta x_B}}{\Delta x_B}\right]\left(\Delta x_B\right)^2\\
& \simeq \frac{1}{2}\frac{\partial^2G_m}{\partial x_B^2}\left(\Delta x_B\right)^2
\end{split}
\end{equation}

If only the bulk thermodynamics is considered this mean that if the average composition is inside the spinodal then \emph{any} fluctuation in composition would result in a reduction in Gibbs energy;  \emph{any} fluctuation in composition would grow. This is not observed experimentally, which mean that there is some energy barrier that must be overcome.

\subsubsection{The gradient energy for a nearest neighbor model}

\begin{figure}
\begin{center}
\includegraphics[width=0.9\textwidth]{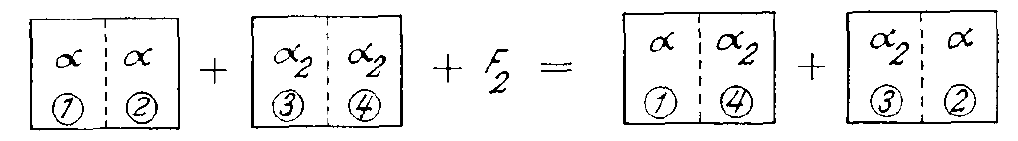}
\end{center}
\caption{Abbildung 6 from Becker \cite{1938bec} showing his thought experiment where two homogeneous alloys $\alpha$ and $\alpha_2$ are cut and put together with a counterpart. $F_2$ is the resulting change in bond energy.}\label{fig:becker}
\end{figure}

Becker in 1938 \cite{1938bec} did a simple thought experiment. He considered two pieces of homogeneous binary alloys $A-B$ consisting of the same phase but having different composition. The hypothetical alloys were cut in two and put together again coherently with the alloy of the differing composition, see Fig.~\ref{fig:becker}. Becker did not consider stresses and only considered nearest neighbor bond energies $E_{AA}$, $E_{BB}$ and $E_{AB}$. He found that, due to the introduced composition gradient, the total bond energy had changed by an amount $\Delta E$ equal to 

\begin{equation}
\Delta E = zV_{AB}\left(\Delta x_B\right)^2
\end{equation}

where $z$, for a given atom,  is the number of nearest neighbors on a neighboring plane, $V_{AB}$ is the so-called interaction energy and $\Delta x_B$  is the difference in composition between the two alloys. The interaction energy $V_{AB}$  is given by

\begin{equation}
V_{AB}=E_{AB}-\frac{1}{2}\left(E_{AA}+E_{BB}\right)
\end{equation}

It is clear that this \emph{gradient energy} derived by Becker will constitute an energy barrier against spinodal decomposition if $V_{AB}>0$, which is always the case; For a nearest neighbor model the interaction energy $V_{AB}$ is directly proportional to the regular solution parameter $L_{AB}$ and has the same sign.

\subsubsection{Gibbs energy of a single phase system\\ with concentration gradients}

Cahn and Hilliard \cite{1958cah}, inspired by Hillert \cite{1956hil}\cite{1961hil}, performed a mathematically elegant derivation of the energy of a nonuniform binary system. They wrote the energy as a Taylor series in which the composition and its spatial derivatives are treated as independent variables. From symmetry considerations and by keeping only the leading terms, the total energy of a single phase binary system can be written as the functional

\begin{equation}\label{eq:gibbsfunc}
G=\int g\,dV=\int \left[g_0+\kappa_{AB}\vert\nabla c_B\vert^2\right]dV
\end{equation}

where $g$ is Gibbs energy per unit volume, $g=G/V$, $c_B$ is concentration of $B$, $c_B=N_B/V$, $\kappa_{AB}$ is a gradient energy coefficient related to the gradient energy found by Becker and $g_0$ is Gibbs energy for a homogeneous material, i.e.~in the absence of concentration gradients. For a regular A-B solution the relation between $\kappa$ and the regular solution parameter $L_{AB}$ is

\begin{equation}
\kappa=\frac{b^2}{2}L_{AB}V_m
\end{equation}

where $b$ is the interplanar distance and $V_m$ the molar volume. 

\subsubsection{Dynamics of spinodal decomposition}

If gradient energy effects are disregarded, the driving force for diffusion is a spatial gradient in chemical potential $\mu_B=\partial G/\partial N_B$. When gradient energy is considered and Gibbs energy therefore is given by the functional Eq.~\ref{eq:gibbsfunc}, which, specifically, contain a term which depend on the gradient in composition the quantity corresponding to $\mu_B$ is given by the functional derivative

\begin{equation}
\frac{\delta G}{\delta c_B}=\frac{\partial g}{\partial c_B}-\nabla\cdot\frac{\partial g}{\partial\left(\nabla c_B\right)}
\end{equation}

Assuming a constant $\kappa$  the final expression is

\begin{equation}\label{eq:funcder}
\frac{\delta G}{\delta c_B}=\mu_B-2\kappa\nabla^2c_B
\end{equation}

It is seen that $-2\kappa\nabla^2c_B$ can be considered as a gradient energy contribution to the chemical potential.

In the absence of gradient energy effects the flux of element $B$ is commonly written as

\begin{equation}
J_B=-M_Bc_B\nabla\mu_B
\end{equation}

where $M_B$ is the mobility of $B$. In order to study the dynamics of spinodal decomposition the flux expression need to be written as

\begin{equation}\label{eq:fluxgrad}
J_B=-M_Bc_B\nabla\left(\mu_B-2\kappa\nabla^2c_B\right)
\end{equation}

which can be combined with the equation of continuity

\begin{equation}\label{eq:cont}
\frac{\partial c_B}{\partial t}=\nabla\cdot\left(-J_B\right)
\end{equation}

to obtain

\begin{equation}\label{eq:spindecfinal}
\frac{\partial c_B}{\partial t}=\nabla\cdot\left[M_Bc_B\nabla\left(\mu_B-2\kappa\nabla^2c_B\right)\right]
\end{equation}

Given an initial state Eq.~\ref{eq:spindecfinal} can be solved to yield the dynamics of spinodal decomposition. This was first done by Cahn \cite{1961cah}. Eq.~\ref{eq:spindecfinal} is often referred to as the Cahn--Hilliard equation.

It should be noted that the effect of stresses due to concentration gradients is not included in Eq.~\ref{eq:spindecfinal}. Coherency stress will act to suppress the spinodal decomposition. This is also discussed by Cahn \cite{1961cah}. 

\subsection{The concept of diffuse interfaces}

\begin{figure}
\begin{center}
\includegraphics[width=1.0\textwidth]{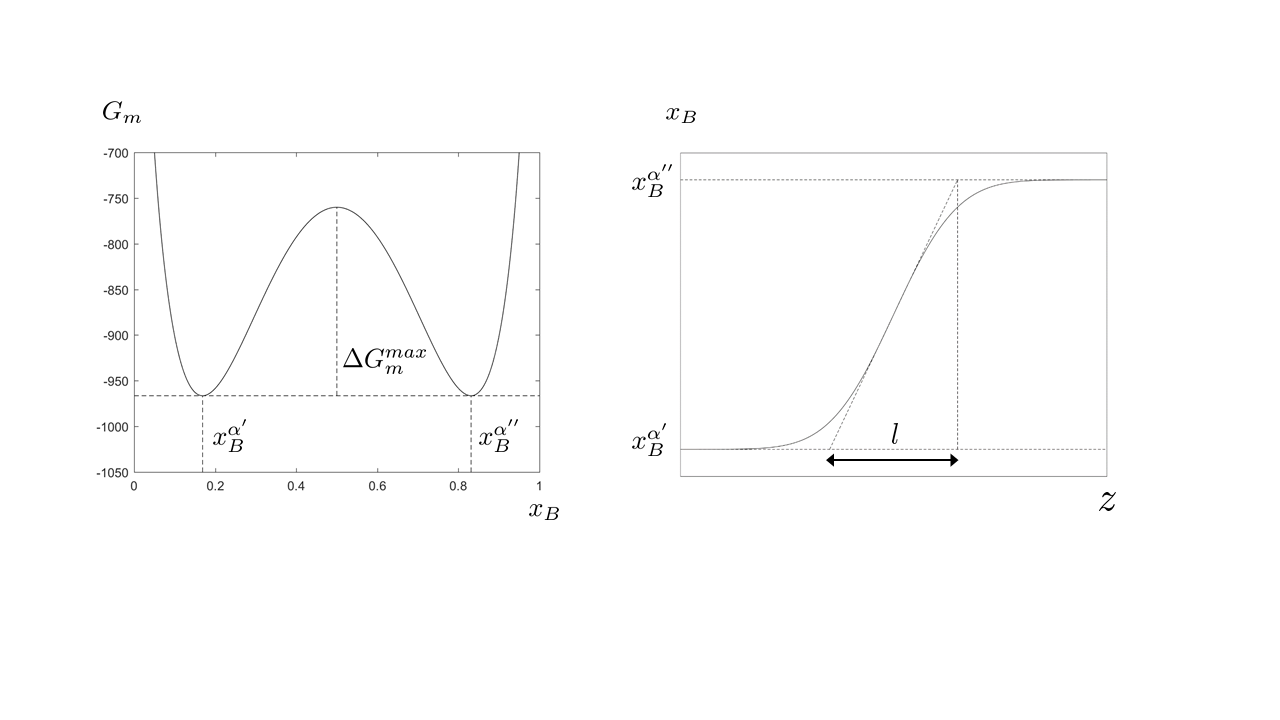}
\end{center}
\caption{The graph to the right show a sketch of a diffuse interface whose approximate width $l$ is given by Eq.~\ref{eq:intwidth}. The graph to the left show $G_m^\alpha$ as a function of composition. The equilibrium compositions $x_B^{\alpha^\prime}$ and $x_B^{\alpha^{\prime\prime}}$ are indicated. After Ref.~\cite{1958cah}.}\label{fig:interface}
\end{figure}

In a spinodally decomposed structure the value of $\kappa$ together with the thermodynamic description determines the width of the transition zone between $\alpha^\prime$ and $\alpha^{\prime\prime}$. According to Cahn and Hilliard \cite{1958cah} the width $l$ of the interface is approximately

\begin{equation}\label{eq:intwidth}
l\simeq\left(x_B^{\alpha^{\prime\prime}}-x_B^{\alpha^{\prime}}\right)\sqrt{\frac{\kappa}{\Delta G_m^{max}}}
\end{equation}

where $x_B^{\alpha^{\prime\prime}}$ and $x_B^{\alpha^{\prime}}$, the equilibrium compositions, and $\Delta G_m^{max}$ are indicated in Fig.~\ref{fig:interface}.

However, the transition between $\alpha^{\prime}$ and $\alpha^{\prime\prime}$ is gradual and the interface is therefore said to be diffuse, see Fig.~\ref{fig:interface}. This is a central concept of the phase-field method, but note that the interface in a spinodally decomposed structure separates regions that differ only in composition (if a single crystal grain is considered), not in structure or orientation.

\subsection{Migration of interfaces}

\subsubsection{Migration of antiphase boundaries}

In Ref.~\cite{1979all} Allen and Cahn took the diffuse interface concept further by considering other types of interfaces. They focused specifically on the motion of antiphase boundaries (APBs).  They wrote the total Gibbs energy of the system as the functional

\begin{equation}\label{eq:funcorder}
G=\int g\, dV=\int\left[g_0+\sigma\vert\nabla\eta\vert^2\right]dV
\end{equation}

where $\eta$ is a parameter describing the local degree of ordering. Gibbs energy as a function of $\eta$ is symmetric around $\eta =0$, i.e. $G(\eta )=G(-\eta)$ and has minima at $\pm\eta_e$. APBs thus occur where there is a transition from $-\eta_e$ to $+\eta_e$. The integrated energy across such a transition constitute the APB energy.

Eq.~\ref{eq:funcorder} is clearly very similar to Eq.~\ref{eq:gibbsfunc} and $\sigma$ is a gradient energy coefficient just as $\kappa$. The local corresponding potential of $G$ is again given by a functional derivative similar to Eq.~\ref{eq:funcder}, viz.

\begin{equation}
\frac{\delta G}{\delta\eta}=\frac{\partial g}{\partial\eta}-\nabla\cdot\frac{\partial g}{\partial\left(\nabla\eta\right)}
\end{equation}

 However, when solving for the evolution in time of $\eta$, as compared to solving for $c_B$, there is a fundamental difference in that $\eta$ is not conserved; the number of moles of $B$ in a closed system is constant, but the volume fraction of the system having one or the other equilibrium value of $\eta$ will generally change over time. Allen and Cahn suggested that the temporal evolution of $\eta$ be given by
 
 \begin{equation}
 \frac{\partial\eta}{\partial t}=-M_{\eta}\frac{\delta G}{\delta\eta}
 \end{equation}
 
 as opposed to the equation of continuity, Eq.~\ref{eq:cont}, used to solve for the temporal evolution of $c_B$. $M_\eta$ is a kinetic coefficient yielding the mobility of the APB.
 
 Given an initial state, the temporal and spatial evolution of $\eta$ results in a monotonic decrease in APB energy, under isothermal conditions.  
 
 \subsection{Migration of phase interfaces}
 
 The treatment by Allen and Cahn is essentially directly applicable to the migration of phase interfaces and grain boundaries. Collins and Levine \cite{1985col} applied it to solidification of a pure substance including thermal diffusion.
 
  \subsection{Anisotropic surface energy}
 
 Anisotropic surface energy is an important factor with regards to the evolution of microstructural morphology. Kobayashi \cite{1993kob} considered this when simulating dendritic crystal growth. To incorporate this in a phase field formulation, a gradient energy coefficient $\varepsilon$ was taken to be a function of the gradient of a phase field variable $\phi$, i.e.~$\varepsilon\left(\nabla\phi\right)$.
 
 \begin{equation}
 G=\int g\,dV=\int\left[g_0+\varepsilon\vert\nabla\phi\vert^2\right]dV
 \end{equation}
 
  \begin{equation}
 \frac{\delta G}{\delta\phi}=\frac{\partial g_0}{\partial\phi}-\nabla\cdot\left[\frac{\partial\varepsilon}{\partial\left(\nabla\phi\right)}\vert\nabla\phi\vert^2+2\varepsilon\nabla\phi\right]
 \end{equation}
 
 $\phi$ take the value one in the solid and zero in the liquid, $0\le\phi\le 1$. See also section \ref{sec:anis}.
 
%--------------

\section{The YAPFI phase-field implementation}\label{sec:pfmodel}

The basis of the Yapfi implementation is the so-called WBM model (Wheeler, Boettinger and McFadden) \cite{1992whe}\cite{1993whe}.

\subsection{Formal description}

The Gibbs energy functional is written as

\begin{equation}
\begin{split}
G &=\int g\, dV\\
&=\int\left[g_0+\sum\nabla c_i\cdot\sum\nabla c_j\kappa_{ij}+\sum\nabla \phi_i\cdot\sum\nabla\phi_j\varepsilon_{ij}\right]dV
\end{split}
\end{equation}

\begin{equation}
\varepsilon_{ii}=\kappa_{ii}=0
\end{equation}

\begin{equation}
\varepsilon_{ij}=\varepsilon_{ji}
\end{equation}

\begin{equation}
\kappa_{ij}=\kappa_{ji}
\end{equation}

where $g$ is Gibbs energy per unit volume, $[J/m^3]$, $c_j$ is concentration of component $j$, $[mol/m^3]$. The $\phi_i$ are phase field variables, $0\le\phi_i\le 1$. In general, a specific phase-field variable is associated with a specific phase. Different phase-field variables associated with the same phase generally represent different crystal grains.

In general, $\varepsilon_{ij}<0$.

For convenience we let

\begin{equation}
g_0=g_1+g_2
\end{equation}

where $g_1$ is a function of the bulk Gibbs energies of the participating phases and the phase-field variables and $g_2$ is a function of composition and the phase-field variables. The derivative $\partial g_1/\partial c_k$ will then be a function of the bulk chemical potentials of component $k$. The function $g_2$ is associated with the interfacial energy and can also be used to introduce Gibbs energy contributions due to specific coupling effects between composition and phase-field variables. Thus,

\begin{align}
g_1& =\sum_rg^rp^r\left(\boldsymbol{\phi}\right)\\
\frac{\partial g_1}{\partial c_k}& =\sum_r\mu_k^r\, p^r\left(\boldsymbol{\phi}\right)\\
g_2& =f\left(\mathbf{c},\boldsymbol{\phi}\right)
\end{align}

where $r$ is a phase index and $p^r\left(\boldsymbol{\phi}\right)$ is an interpolating polynomial. Though $g_1$ is written as a sum over the participating phases, since that is the common form, arbitrary expressions are allowed. 

The temporal evolution of composition variable $c_j$ is given by a slightly modified Cahn--Hilliard equation \cite{1961cah} in order to take into account the effect of a temperature gradient (see also section \ref{sec:homo})

\begin{equation}
 \frac{\partial c_j}{\partial t} =\nabla\cdot\Bigg\lbrace M_jc_j\left[\nabla\left(\frac{\delta G}{\delta c_j}\right)+\frac{Q_j^\star}{T}\nabla T\right]\Bigg\rbrace
\end{equation}

The parameter $Q_j^\star$ is the so-called heat of transport \cite{1979nic}.

The temporal evolution of phase-field variable $\phi_k$ is given by the Allen--Cahn equation \cite{1979all}

\begin{equation}
\frac{\partial\phi_k}{\partial t}=-M_\phi\frac{\delta G}{\delta\phi_k}
\end{equation}

With $g_0(\mathbf{c},\boldsymbol\phi,T)$, $\kappa_{ij}(\mathbf{c},\boldsymbol\phi,T)$ and  $\varepsilon_{ij}(\mathbf{c},\boldsymbol\phi,\nabla\boldsymbol\phi,T)$ --- arbitrary analytical expressions are allowed for $g_0$, $\boldsymbol{\kappa}$ and $\boldsymbol{\varepsilon}$ --- the variational derivatives are evaluated as

\begin{equation}\label{eq:vardgdc}
\begin{split}
\frac{\delta G}{\delta c_k} &=\frac{\partial g}{\partial c_k}-\nabla\cdot\frac{\partial g}{\partial\left(\nabla c_k\right)}\\
&=\frac{\partial g_1}{\partial c_k}+\frac{\partial g_2}{\partial c_k}+\\
&\sum\nabla c_i\cdot\sum\nabla c_j\frac{\partial\kappa_{ij}}{\partial c_k}+\sum\nabla\phi_i\cdot\sum\nabla\phi_j\frac{\partial\varepsilon_{ij}}{\partial c_k}\\
&\quad -\nabla\cdot\left(2\sum\nabla c_j\kappa_{jk}\right)\\
&=\sum_r\mu_k^r\, p^r\left(\phi\right)+\frac{\partial g_2}{\partial c_k}+\\
&\sum\nabla c_i\cdot\sum\nabla c_j\frac{\partial\kappa_{ij}}{\partial c_k}+\sum\nabla\phi_i\cdot\sum\nabla\phi_j\frac{\partial\varepsilon_{ij}}{\partial c_k}\\
&\quad -2\sum\left(\nabla^2c_j\kappa_{jk}+\nabla c_j\cdot\nabla\kappa_{jk}\right)
\end{split}
\end{equation}

\begin{equation}
\begin{split}
\frac{\delta G}{\delta\phi_k} &=\frac{\partial g}{\partial\phi_k}-\nabla\cdot\frac{\partial g}{\partial\left(\nabla\phi_k\right)}\\
&=\frac{\partial g_1}{\partial\phi_k}+\frac{\partial g_2}{\partial\phi_k}+\\
&\sum\nabla c_i\cdot\sum\nabla c_j\frac{\partial\kappa_{ij}}{\partial\phi_k}+\sum\nabla\phi_i\cdot\sum\nabla\phi_j\frac{\partial\varepsilon_{ij}}{\partial\phi_k}\\
&\quad -\nabla\cdot\left(2\sum\nabla\phi_j\varepsilon_{jk}\right)-\nabla\cdot\left(\sum\nabla\phi_i\cdot\sum\nabla\phi_j\frac{\partial\varepsilon_{ij}}{\partial\left(\nabla\phi_k\right)}\right)\\
&=\frac{\partial g_1}{\partial\phi_k}+\frac{\partial g_2}{\partial\phi_k}+\\
&\sum\nabla c_i\cdot\sum\nabla c_j\frac{\partial\kappa_{ij}}{\partial\phi_k}+\sum\nabla\phi_i\cdot\sum\nabla\phi_j\frac{\partial\varepsilon_{ij}}{\partial\phi_k}\\
&\quad-2\sum\left(\nabla^2\phi_j\varepsilon_{jk}+\nabla\phi_j\cdot\nabla\varepsilon_{jk}\right)-\nabla\cdot\boldsymbol{\xi}_k\\
\end{split}
\end{equation}

where the vector quantity $\boldsymbol{\xi}_k=[\xi_{kx}\, \xi_{ky}\, \xi_{kz}]$  was introduced for convenience

\begin{equation}
\boldsymbol{\xi}_k=\sum\nabla\phi_i\cdot\sum\nabla\phi_j\frac{\partial\varepsilon_{ij}}{\partial\left(\nabla\phi_k\right)}
\end{equation}

It should be noted that when taking the derivatives $\partial /\partial(\nabla c_k)$ and $\partial /\partial(\nabla\phi_k)$ the result is a vector, e.g. 

\begin{equation*}
\partial\varepsilon_{ij}/\partial (\nabla\phi_k)=\Big[\partial\varepsilon_{ij}/\partial (\partial\phi_k/\partial x)\quad\partial\varepsilon_{ij}/\partial (\partial\phi_k/\partial y)\quad\partial\varepsilon_{ij}/\partial (\partial\phi_k/\partial z)\Big]
\end{equation*}

\begin{equation*}
\begin{split}
\sum_i\nabla\phi_i\cdot\sum_j \nabla\phi_j\,\partial\varepsilon_{ij}/\partial (\nabla\phi_k)= &\Big[\sum_i\nabla\phi_i\cdot\sum_j \nabla\phi_j\,\partial\varepsilon_{ij}/\partial (\partial\phi_k/\partial x)\\
&\sum_i\nabla\phi_i\cdot\sum_j \nabla\phi_j\,\partial\varepsilon_{ij}/\partial (\partial\phi_k/\partial y)\\
&\sum_i\nabla\phi_i\cdot\sum_j \nabla\phi_j\,\partial\varepsilon_{ij}/\partial (\partial\phi_k/\partial z)\Big]
\end{split}
\end{equation*}

$\nabla\cdot (\sum_i\nabla\phi_i\sum_j\cdot \nabla\phi_j[\partial\varepsilon_{ij}/\partial (\nabla\phi_k)])$ is scalar.

All derivatives are evaluated analytically.

Fourier's law yields the heat flux $J_q$

\begin{equation}
J_q=-\lambda\nabla T\quad\quad\left[J\cdot m^{-2}\cdot s^{-1}\right]
\end{equation}

where $\lambda$ is the thermal conductivity.

The heat flux is combined with a conservation law

\begin{equation}
\frac{\partial Q}{\partial t}=\nabla\cdot\left(-J_q\right)+\dot{q}\quad\quad\left[J\cdot m^{-3}\cdot s^{-1}\right]
\end{equation}

where $\dot{q}$ is a source term. If only isobaric conditions are considered $Q$ equals volumetric enthalpy.

The relation between $\partial Q /\partial t$ and $\partial T /\partial t$ is

\begin{equation}
\frac{\partial Q}{\partial t}=\frac{c_p}{V_m}\dot{T}
\end{equation}

The source term $\dot{q}$ is given by

\begin{equation}
\dot{q}=\frac{\sum_r H_m^r\dot{\phi}_r}{V_m}=\sum_r H_V^r \dot{\phi}_r
\end{equation}

The final expression is

\begin{equation}
\left(\sum_r\frac{c_p^r}{V_m}\phi_r\right)\dot{T}=\nabla\cdot\left(\lambda\nabla T\right)+\sum_r\frac{H_m^r}{V_m}\dot{\phi}_r
\end{equation}

\subsection{Notes on input data and implementation}

In the formal description above concentration $c_k$ $[mol/m^3]$ is used, but internally the software is mainly using mole fractions $x_k$, which is reflected in the required format of input data. 

When entering data, the function $g_1$ is allowed to be an arbitrary function of the molar Gibbs energies of the participating phases $G_m^\alpha$ and the phase-field variables $\phi_k$. The $G_m^\alpha$ $[J/mol]$ should, in principle, be valid for a microscopically homogeneous single phase system.

The function $g_2$ is allowed to be an arbitrary function of R, T the mole fractions of elements $x_j$ and the phase-field variables $\phi_k$. As $g_1$, the input unit of $g_2$ is $[J/mol]$.

The gradient energy coefficients $\varepsilon_{ij}$ and $\kappa_{ij}$ are allowed to be arbitrary functions of R, T the mole fractions of element $x_j$ and the phase-field variables $\phi_k$. In addition, $\varepsilon_{ij}$ may be a function of the gradients of the phase-field variables $\nabla\phi_k$, though in general the function should then be formulated such that it is a function of the normal of phase-field variables $\hat{n}_k=\nabla\phi_k/\vert\nabla\phi_k\vert$.

The unit of input functions $\varepsilon_{ij}$ and $\kappa_{ij}$ is  $[J\cdot m^2\cdot mol^{-1}]$.

$M_\phi$ should be input with unit $[mol\cdot J^{-1}\cdot s^{-1}]$ whereas $M_k$ has unit $[m^2\cdot mol\cdot J^{-1}\cdot s^{-1}]$. In the formal description the unit of $M_\phi$ is $[m^3\cdot J^{-1}\cdot s^{-1}]$, i.e.~it differs by a factor $1/V_m$ from the required unit of input data.

With the exception of analytical expressions entered for $G_m$, the interdependece of $x_i$, $\phi_i$ and $\nabla\phi_i$, respectively, is taken into account when evaluating partial derivatives with respect to these variables. The exception made for $G_m$ is due to the fact that this is not necessary when evaluating chemical potentials from $G_m$ and $\partial G_m/\partial x_i$. 

\subsection{Interfacial energy}

The total interfacial energy of a system $E^{int}$ $[J]$ is typically given by

\begin{equation}
E^{int}=\int\frac{1}{V_m}\left(g_2^\prime+\sum\nabla\phi_i\cdot\sum\nabla\phi_j\varepsilon_{ij}^\prime\right) dV
\end{equation}

The ``primes'' have been added to indicate that the units here are those required in the input, i.e.~$[J/mol]$ and $[J\cdot m^2\cdot mol^{-1}]$ for $g_2^\prime$ and $\varepsilon_{ij}^\prime$, respectively.

\section{Homogenization model}\label{sec:homo}

The so-called homogenization model \cite{2006lar}\cite{2009lar} is included in the implementation. This model by itself is intended for multiphase simulations in which it is assumed that the material is always fully locally equilibrated with respect to phase fractions, phase compositions etc.~and thus do not make use of phase field variables and do not allow for any supersaturation with respect to the phases entered in a simulation. This model is suitable for simulations where diffusion distances are considerably larger than some characteristic microstructural length scale, such as the interparticle spacing or grain size, depending on the type of simulation. The model traces its roots to previous work by, among others, Engstr\"om et al.~\cite{1994eng} and Morral et al.~\cite{1992mor}. 

In a single phase $\alpha$ system the flux of element $k$ is given by

\begin{equation}
J_k^\alpha =-M_k^\alpha c_k^\alpha\nabla\mu_k^\alpha
\end{equation}

With the homogenization model the flux through a multiphase mixture is considered. Let the permeability of element $k$ in phase $\alpha$ be given by

\begin{equation}
\Gamma_k^\alpha =M_k^\alpha c_k^\alpha
\end{equation}

The effective permeability $\Gamma_k^\star$ through a multiphase mixture is then assumed to be given by some averaging procedure, for example a rule of mixtures

\begin{equation}
\Gamma_k^\star =\sum_\beta f^\beta \Gamma_k^\beta
\end{equation}

where the summation is taken over all phases present locally. $f^\beta$ is the local volume fraction of phase $\beta$. Since it is assumed that full local equilibration is maintained, the chemical potentials are obviously the same locally in all phases. The flux through the multiphase mixture is thus given by

\begin{equation}
J_k =-\Gamma_k^\star\nabla\mu_k^{\text{l.eq.}}
\end{equation}

The ``l.eq.'' has been added to emphasize the assumption of local equilibration.

The flux expression is then combined with the equation of continuity

\begin{equation}
\frac{\partial c_k}{\partial t}=\nabla\cdot\left(-J_k\right)
\end{equation}

The homogenization model may be combined with phase-field by letting multiple phases be associated with a single phase-field variable. 

If there is a temperature gradient the flux expression is

\begin{equation}
J_k =-\Gamma_k^\star\left(\nabla\mu_k^{\text{l.eq.}}+\frac{Q_k^\star}{T}\nabla T\right)
\end{equation}

\section{Implementation}

The model was implemented in Fortran using a finite volume approach and allow for 1-3D simulations. The implementation is fully implicit, i.e.~all coefficients are evaluated implicitly. The degree of implicity is arbitrary (e.g.~trapezoidal rule $(\theta =0.5)$ or Euler backward $(\theta =1)$). Arbitrary analytical expressions are supported for $g_0(\mathbf{c},\boldsymbol\phi ,T)$, $\kappa_{ij}(\mathbf{c},\boldsymbol\phi ,T)$ and  $\varepsilon_{ij}(\mathbf{c},\boldsymbol\phi,\nabla\boldsymbol\phi ,T)$. All derivatives are evaluated analytically.

Thermodynamic and kinetic data may be obtained from Calphad type databases using the TQ API \cite{2002and} utilizing a so-called interpolation scheme \cite{2015lar} or from analytical expressions entered for the molar Gibbs energy of participating phases. However, in the latter case phases may not exhibit internal degrees of freedom.

\section{Example simulations}

\subsection{Simple phase transformation in 1D and comparison with a sharp interface (Dictra) simulation}\label{sec:compdictra}

Growth of a phase $\alpha$ from a supersaturated parent $\beta$ phase was considered for a hypothetical binary system. Both phases are regular solutions. A corresponding sharp interface simulation using Dictra \cite{2000bor} was performed. 

For this simulation the $g_2$ phase interface Gibbs energy contribution was given by the method suggested by Finel et al. \cite{2018fin}. That method has the advantage that it allows for very thin phase interfaces, i.e.~extending over very few grid points.

\begin{figure}
\begin{center}
\includegraphics[width=0.8\textwidth]{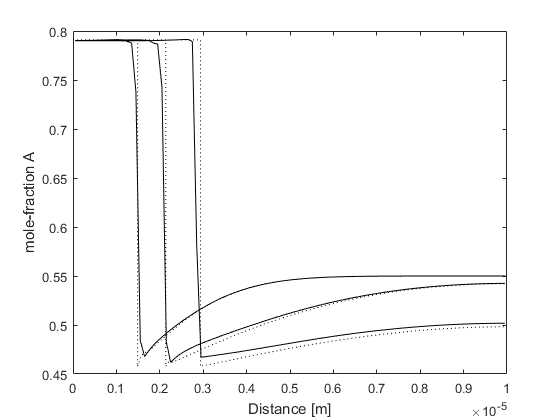}
\end{center}
\caption{Concentration profiles at different times during diffusion controlled growth for a hypothetical binary system. Corresponding profiles obtained from a Dictra simulation show as dotted lines.}\label{fig:compare_dictra}
\end{figure}

Concentration profiles at different times are shown in Fig.~\ref{fig:compare_dictra}. The profiles obtained from Dictra are shown with dotted lines. The agreement is very satisfactory and could of course be even better with a denser grid.

\subsection{Diffusion controlled growth influenced by anisotropic surface energy}\label{sec:anis}

For simplicity only 2D simulations will be considered. Following Kobayashi \cite{1993kob}, let

\begin{equation}
\varepsilon =\bar{\varepsilon}\sigma
\end{equation}

where $\bar{\varepsilon}$ is a mean surface energy and

\begin{equation}\label{eq:sigma}
\sigma\left(\theta\right)=\lbrace 1+\delta\cos\left[j\left(\theta -\theta_0\right)\right]\rbrace^2
\end{equation}

The normal vector at the interface is given by

\begin{equation}
\hat{n}=\frac{\nabla\phi}{\vert\nabla\phi\vert}
\end{equation}

Let the reference direction point in the positive x direction, then, 

\begin{equation}
\theta =\arccos\left(\frac{\displaystyle\frac{\partial\phi}{\partial x}}{\sqrt{\displaystyle\left(\frac{\partial\phi}{\partial x}\right)^2+\displaystyle\left(\frac{\partial\phi}{\partial y}\right)^2}}\right)
\end{equation}

\begin{figure}
\centering

\begin{subfigure}[t]{0.4\textwidth}
\centering
\includegraphics[width=\textwidth]{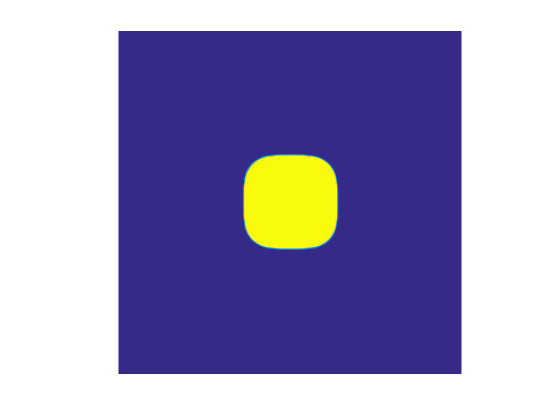}
\caption{$j=4$, $\delta =0$}\label{fig:no_anis}
\end{subfigure}
~
\begin{subfigure}[t]{0.4\textwidth}
\centering
\includegraphics[width=\textwidth]{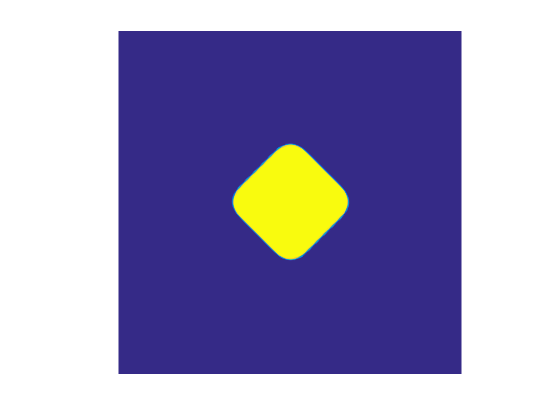}
\caption{$j=4$, $\delta =0.1$}\label{fig:low_anis}
\end{subfigure}
~
\begin{subfigure}[t]{0.4\textwidth}
\centering
\includegraphics[width=\textwidth]{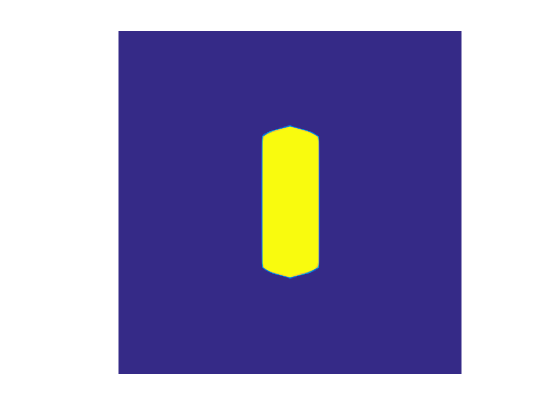}
\caption{$j=6$, $\delta =0.1$}\label{fig:anis_6fold}
\end{subfigure}
~
\begin{subfigure}[t]{0.4\textwidth}
\centering
\includegraphics[width=\textwidth]{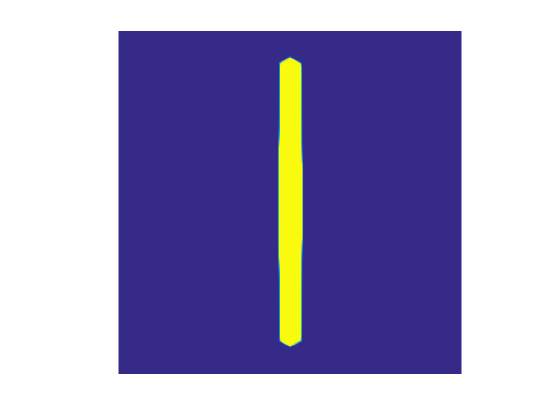}
\caption{$j=6$, $\delta =0.2$}\label{fig:anis_6fold}
\end{subfigure}

\caption{Simulations of diffusion controlled growth where surface energy is anisotropic. Cf.~Eq.~\ref{eq:sigma} for the parameter values. Though Fig.~\ref{fig:no_anis} has no explicitly entered anisotropy the inherent anisotropy of the square numerical grid influence the precipitate shape.}\label{fig:anis}

\end{figure}

Results from simulations using different values of $\delta$ are shown in Fig.~\ref{fig:anis}. The same hypothetical A-B system as in section \ref{sec:compdictra} was considered, and, again, growth of $\alpha$ from supersaturated $\beta$ was simulated.

As stated above, $\varepsilon$ can be an arbitrary analytical function of $\mathbf{c}$, $\boldsymbol{\phi}$ and $\nabla\boldsymbol{\phi}$. When setting up the simulation and entering the expression for $\varepsilon$, the derivatives $\partial\phi /\partial x$ and $\partial\phi /\partial y$ are thus two of the allowed variables.

\subsection{Spinodal decomposition}

\begin{figure}
\centering

\begin{subfigure}[t]{0.5\textwidth}
\centering
\includegraphics[width=\textwidth]{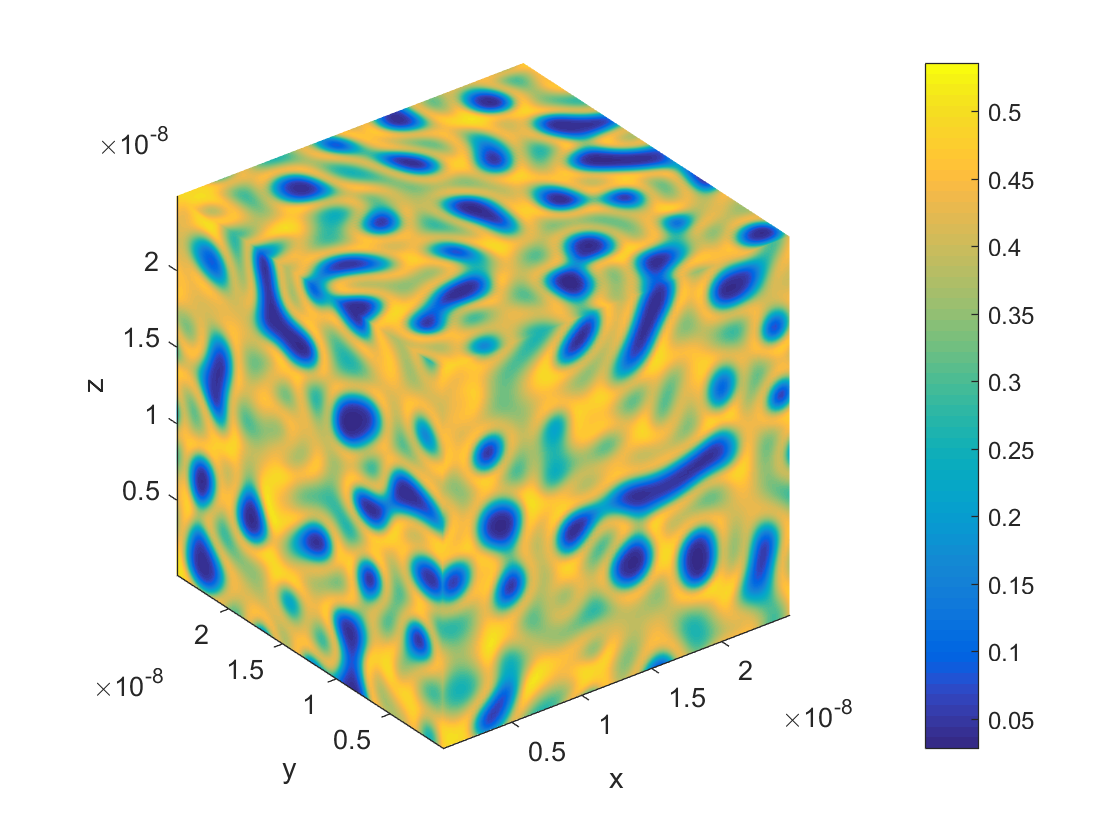}
\caption{Mole fraction of element A.}\label{fig:spinodal3da}
\end{subfigure}
~
\begin{subfigure}[t]{0.5\textwidth}
\centering
\includegraphics[width=\textwidth]{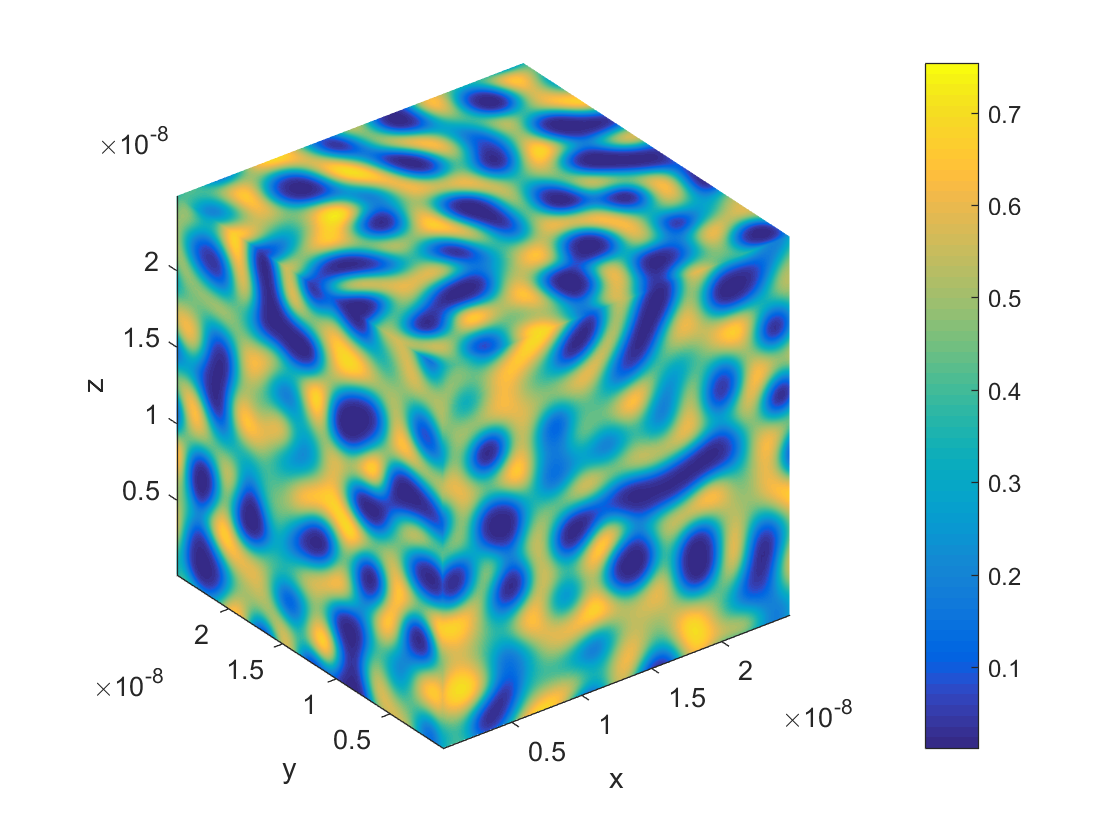}
\caption{Mole fraction of element B.}\label{fig:spinodal3db}
\end{subfigure}

\begin{subfigure}[t]{0.5\textwidth}
\centering
\includegraphics[width=\textwidth]{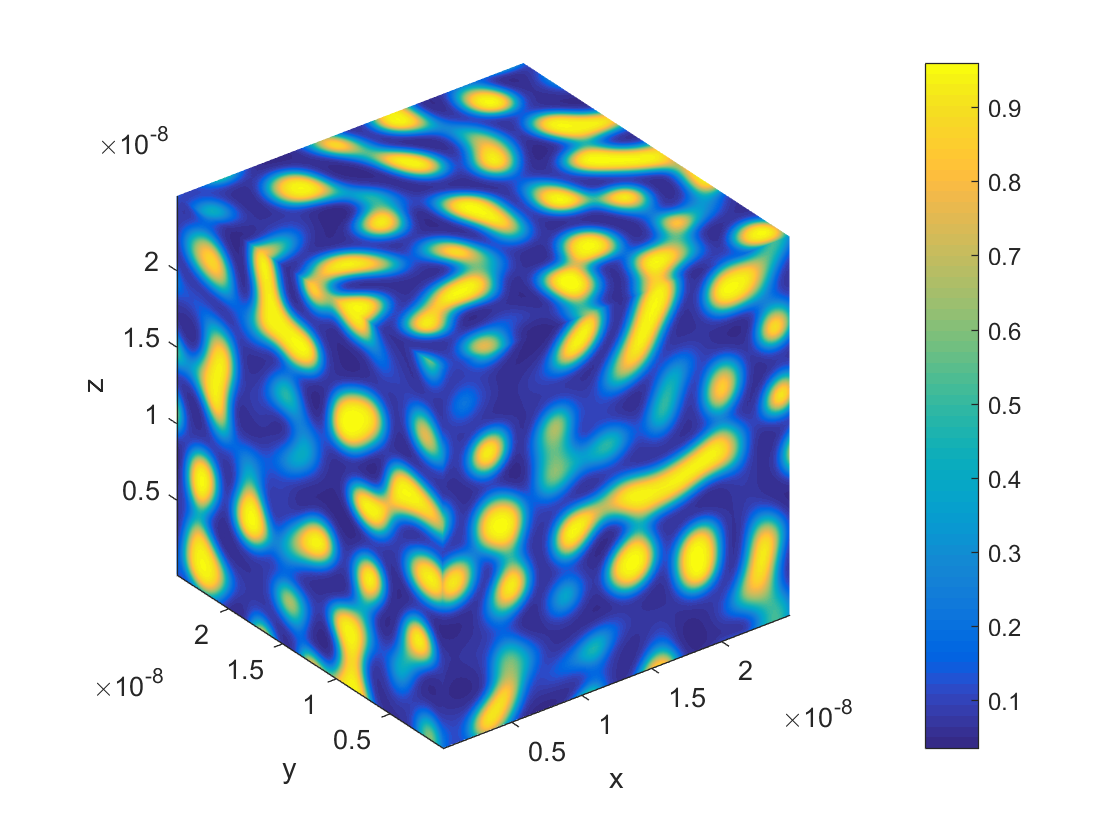}
\caption{Mole fraction of element C.}\label{fig:spinodal3dc}
\end{subfigure}

\caption{Simulation of spinodal decomposition for a ternary system.}\label{fig:spindec3d}

\end{figure}

In Fig.~\ref{fig:spindec3d} are shown results from a 3D simulation of spinodal decomposition. The simulation was performed for a hypothetical A-B-C system consisting of a single regular solution phase in which all regular solution parameters $L_{ij}> 0$, but with different values. The domain size was $\left(25\cdot 10^{-9}\right)^3\,\left[\text{m}^3\right]$ with $100^3$ grid points. Periodic boundary conditions were applied. The gradient energy coefficients were set to $\kappa_{ij}=-d^2L_{ij}/2$ where d is the grid spacing, i.e.~$0.25\,\text{nm}$. Initial mole-fractions were $x_A=x_B=0.33,\, x_C=0.34$ with a random noise in the range  $\pm 0.01$.

\subsection{Homogenization model simulation}

For an example of a 2D simulation using the homogenization model the reader is referred to the work by Salmasi et al.~\cite{2019sal}. In that work geometry effects during gradient sintering of cemented carbides was investigated.

\end{document}